\documentclass[11pt]{article}
\usepackage{geometry}
\usepackage{fancyhdr}
\usepackage{graphicx}
\usepackage{graphics}
\usepackage{amsmath}
\usepackage{amssymb}

\begin{document}

\begin{center}
{\Large\bf Particle acceleration at relativistic shock waves}
\vskip 0.5cm

Martin Lemoine$^1$, Guy Pelletier$^2$
\vskip 0.3cm

{\it $^1$ Institut d'Astrophysique de Paris, CNRS, UPMC, 98 bis
    boulevard Arago, 75014 Paris, France \\
$^2$ Institut de Plan\'etologie et d'Astrophysique de Grenoble,
    CNRS, Universit\'e Joseph Fourier-II, 34 rue de la Piscine, 38041
    Grenoble, France}\\
\end{center}
\vskip 2cm

{\small\noindent{\bf Abstract}.
  Relativistic sources, e.g. gamma-ray bursts, pulsar wind nebulae and
  powerful active galactic nuclei produce relativistic outflows that
  lead to the formation of collisionless shock waves, where particle
  acceleration is thought to take place. Our understanding of
  relativistic shock acceleration has improved in the past decade,
  thanks to the combination of analytical studies and high level
  numerical simulations. In ultra-relativistic shocks, particle
  acceleration is made difficult by the generically transverse
  magnetic field and large advection speed of the shocked plasma. Fast
  growing microturbulence is thus needed to make the Fermi process
  operative. It is thought, and numerical simulations support that
  view, that the penetration of supra-thermal particles in the shock
  precursor generates a magnetic turbulence which in turn produces the
  scattering process needed for particle acceleration through the
  Fermi mechanism. Through the comparison of the growth timescale of
  the microinstabilities in the shock precursor and the precursor
  crossing timescale, it is possible to delimit in terms of
  magnetization and shock Lorentz factor the region in which
  micro-turbulence may be excited, hence whether and how Fermi
  acceleration is triggered.  These findings are summarized here and
  astrophysical consequences are drawn.
}
\vskip 0.5cm

\vskip 2cm

\section{Introduction}
Broadband spectra of powerful astrophysical objects, from the radio
range, through X-rays, gamma-rays up to TeV energies, unerringly
reveal non-thermal powerlaws that are interpreted as the secondary
radiation of accelerated charged particles. Relativistic collisionless
shock waves are most commonly discussed as natural sites for particle
acceleration, for good and various reasons. First and foremost, these
shock waves are the natural consequences of the relativistic outflows
associated with relativistic sources such as microquasars, active
galactic nuclei, gamma-ray bursts, pulsar wind nebulae, which all
harbour a central object producing a relativistic wind. As viewed in
the frame in which the shock front remains at rest, the incoming
kinetic energy flux $\gamma_{\rm sh}^2n_{\rm u}mc^2$ of a cold
upstream population -- $\gamma_{\rm sh}$ denotes the Lorentz factor of
the shock wave, $n_{\rm u}$ the proper density of the upstream
(unshocked) plasma -- flows through the shock and gets converted into
thermal disordered energy with mean energy per particle
$\sim\gamma_{\rm sh}m c^2$. Provided a reasonable fraction ($\sim
1-10$\%) of this energy flux is converted into a non-thermal radiation
powerlaw, one is able to explain the large amount of radiation
produced in these sources. In the particular case of gamma-ray bursts,
there exists a substantial body of evidence arguing in favor of shock
acceleration of electrons at the external shock front of the wind as
it impinges on the circumburst medium in relation to the observed
afterglow emission (e.g. Piran 2005 for a review). There exists also
overwhelming evidence for particle acceleration at the
non-relativistic counterparts of such collisionless shock waves,
either in the interplanetary medium or in supernovae
remnants. Finally, from a theorist perspective, understanding how
particle acceleration takes place around a shock wave appears as a
well posed problem, the physics of which appears controlled by a
limited set of parameters, most notably the Lorentz factor of the
shock wave $\gamma_{\rm sh}$ and the magnetization level
$\sigma$\footnote{For definiteness, we recall that at normal
  incidence, a strong weakly magnetized shock moves at velocity $c/3$
  away from the shocked (downstream) plasma, which itself moves toward
  the unshocked (upstream) plasma with bulk Lorentz factor
  $\gamma_{\rm sh}/\sqrt{2}$ (Blandford \& McKee 1976). The
  magnetization level is defined as the ratio of the incoming magnetic
  energy flux to the incoming kinetic energy flux, $\sigma = B_{\rm
    u}^2/(4\pi n_{\rm u}m_pc^2)$ as measured in terms of upstream
  quantities, i.e. $\sigma = (v_{\rm A}/c)^2$ in terms of the upstream
  Alfv\'en speed (for $\sigma\ll1$). }.

The Fermi process of shock acceleration is well known, at least in the
non-relativistic regime and in the test particle limit in which one
neglects the influence of the accelerated particles on the shock
environment. The incoming plasma flow (as viewed from the shock front
rest frame) is supposed to carry a frozen in turbulent magnetic field;
high energy particles scatter off magnetic perturbations with a mean
free path much larger than the shock thickness, so that they can cross
the shock front back and forth; these particles undergo elastic
interactions with magnetic disturbances in the proper frame of these
latter, however upstream magnetic perturbations move faster than
downstream ones, therefore a particle that undergoes a Fermi cycle --
i.e. a cycle upstream-downstream-upstream or
downstream-upstream-downstream -- picks up energy in a systematic way
from the convective electric fields. Even in this test particle limit,
relativistic shock acceleration reveals substantial differences with
its non-relativistic counterpart, as revealed by Achterberg et
al. (2001). Yet as we have learned in the past ten years, the test
particle limit obliviates effects that are crucial in the relativistic
regime. What is unusual in astrophysics, it becomes mandatory to
consider the microphysics of the acceleration process, in particular
the intimate non-linear relationship between the accelerated particles
and the shock structure, as will be stressed in the following.

This short review is organized as follows. Section 2 discusses first
the test particle picture of relativistic shock acceleration and its
limitation; a more modern view of this process is then
presented. Section 3 discusses some applications to high energy
astrophysics, regarding notably the acceleration in gamma-ray bursts,
pulsar wind nebulae and active galactic nuclei, as well as the
possibility of accelerating particles to ultra-high energies in
relativistic shocks.

\section{Relativistic shock acceleration}

\subsection{A simplified view: the test particle limit}

One of the most dramatic success of Fermi acceleration at shock waves
is to predict the formation of a powerlaw of accelerated particles
with differential energy spectrum index $s\approx 2$,
i.e. approximately constant energy per logarithmic energy bin, thanks
to the competition between energy gain and escape, in good qualitative
agreement with a variety of observed powerlaws. In the
non-relativistic limit, this competition leads to the formula for
diffusive shock acceleration (Bell 1978): $s=1 - {\rm ln}(P_{\rm
  ret})/{\rm ln}(1+\Delta E/E)$, where $P_{\rm ret}$ represents the
probability of remaining at the acceleration site after a Fermi cycle
around the shock front\footnote{for a steady planar shock, escape
  occurs through advection towards downstream with probability
  $1-P_{\rm ret}$, as the shock front moves away from downstream but
  towards upstream} and $\Delta E/E$ the relative energy gain per
cycle. A relativistic variant of this formula has been proposed by
Vietri (2003).

As noted by Gallant \& Achterberg (1999), Achterberg et al. (2001),
relativistic shock acceleration differs substantially from its
non-relativistic counterpart, because the shock wave moves with a
velocity $\sim c$ close to that of the accelerated particle. One of
the most dramatic consequences is to limit the energy gain per cycle
to a modest $\Delta E/E \sim {\cal O}(1)$. Together with an escape
probability per cycle $1-P_{\rm ret}\sim 0.4$, this leads to the
formation of a powerlaw with spectral index $s\sim 2.2-2.3$, at least
if the scattering process is isotropic. This value has been derived
through semi-analytical techniques (Kirk et al. 2000, Achterberg et
al. 2001), Monte Carlo simulations (Bednarz \& Ostrowski 1998, Ellison
\& Double 2002, Lemoine \& Pelletier 2003) and analytical estimates
(Keshet \& Waxman 2005) but one must emphasize that this value assumes
efficient acceleration (to be discussed further below) and isotropic
scattering. The latter might turn out to be a poor assumption, all the
more so as the shock normal sets a privileged direction, e.g.  shock
compression compresses the magnetized turbulence along this
direction. Accounting for such anisotropy, one finds different spectra
(Keshet 2006, Lemoine \& Revenu 2006).

More importantly, if one adheres to the test particle limit, one is
led to the conclusion that Fermi acceleration fails in the highly
relativistic regime unless very special circumstances are met. This
can be argued as follows but first of all, let us stress that
relativistic shock waves are quite generically superluminal (Begelman
\& Kirk 1990): this occurs when the angle $\Theta_B$ between the
magnetic field in the upstream rest frame and the shock normal
verifies $\Theta_B\gtrsim 1/\gamma_{\rm sh}$ if $\gamma_{\rm sh}\gg1$.
Superluminal regime means that the intersection point of a magnetic
field line with the shock surface moves faster than light across this
surface. Hence, if a particle were tied to a field line, it would not
be able to cross repeatedly back and forth the shock front. To put it
otherwise, the magnetic field can be considered as almost
perpendicular in the front frame, since its transverse component is
amplified by a factor $\sim \gamma_{\rm sh}$ relatively to the
longitudinal component (i.e. along the shock normal). Particles must
therefore diffuse across the field lines in order to complete Fermi
cycles.

In principle, cross-field diffusion is possible in large scale
turbulence -- large means here that most of the power lies at a scale
that is much larger than the typical Larmor radius of the accelerated
particle -- and it had been hoped that such diffusion would allow
particle acceleration at relativistic shock waves. But an usual
turbulent MHD state with a large scale coherence length behaves like
an ordered magnetic field for the suprathermal particles, because
their penetration length upstream remains much shorter than the
coherence length of the turbulence and the expected duration of the
cycle is much smaller than the eddy turn over time. It can thus be
shown analytically that, in such turbulence, a particle cannot execute
more than one and a half Fermi cycle before being dragged away far
downstream (Lemoine et al. 2006). As the particle gets advected away,
it takes a timescale $\sim D_{\parallel}/c^2$ (with $D_\parallel$ the
parallel diffusion coefficient in the turbulence) for the particle
pitch angle to start diffusing; however, by that time, the particle
lies much further away from the shock front than a perpendicular
diffusion length, since $D_\parallel \gg D_\perp$ and the shock front
moves away at large velocity $\simeq c/3$. The particle never catches
up with the shock front, as it moves across the field lines at a much
smaller effective velocity. Such inhibition of Fermi cycles in the
relativistic limit has been observed in test particle Monte Carlo
simulations by Niemiec \& Ostrowski (2006) and Niemiec et al. (2006).

When does relativistic Fermi acceleration then take place? In the
simplest case, one can argue that acceleration is to occur provided
intense short scale turbulence has been excited, on scales
$\lambda_B\lesssim r_{\rm L}$ with intensity $\delta B/B\gg 1$
(Pelletier et al. 2009). Under the conditions discussed in this work,
the micro-turbulence unlocks the particles off the field lines that
would otherwise carry them away from the shock front. When and where
such conditions are met nicely explain the results of Niemiec et
al. (2006). More sophisticated scenarios for acceleration include
e.g., radiative interactions during the Fermi cycles (Derishev et
al. 2003) or magnetic dissipation in the shock transition (Lyubarsky
2003); they will be addressed further below.

That short scale turbulence would be excited in the vicinity of a
collisionless shock should not be regarded as a surprise. In the
absence of collisions, small scale electromagnetic fields are to play
the agents that slow down the flow and dissipate the incoming kinetic
energy. The accelerated particles, as forerunners of the shock are
susceptible of exiciting such small scale turbulence. This is where
one has to abandon the test particle description and consider the
backreaction of these accelerated particles on the shock structure.

There exist interesting connections with observations in this
context. In particular, the standard modeling of gamma-ray burst
afterglow spectra points to a much higher level of magnetized
turbulence in the shocked region than in the interstellar medium, by
some 5 orders of magnitude in magnetic field intensity (Gruzinov \&
Waxman 1999). This implies that the turbulence is self-generated in
the blast. A leading candidate for the source of such turbulence is
the Weibel (filamentation) instability in the shock precursor, between
the accelerated particle population and the incoming (upstream) plasma
population (Medvedev \& Loeb 1999), which produces filaments on a
plasma scale $\sim c/\omega_{\rm pi}$ (with $\omega_{\rm pi}$ the
upstream ion plasma frequency). If the gamma-ray burst explodes in the
interstellar medium, $c/\omega_{\rm pi}\sim 10^7\,$cm is much smaller
than the typical Larmor radius of suprathermal particles $r_{\rm
  L,0}\sim 10^{12}\,$cm (as measured in the downstream frame, in the
shock compressed interstellar field).

Moreover, by comparing the acceleration and the cooling timescale, Li
\& Waxman (2006) [see also Li \& Zhao (2011)] have argued that the
generation of the early X-ray afterglow requires a much larger
magnetic field in the shock precursor than in the interstellar medium.
Unless the circumburst medium is substantially magnetized -- a
possibility considered further below -- turbulence must have been
excited in the precursor, and if so, on very short length
scales. Indeed, the precursor itself is at most of extent $r_{\rm
  L,0}/\gamma_{\rm sh}^3\sim 10^{10}\,$cm, as measured in the upstream
rest frame, with $r_{\rm L,0}$ denoting the Larmor radius of
suprathermal particles in the background magnetic field, the numerical
value corresponding to $B_{\rm u}=1\,\mu$G and $\gamma_{\rm
  sh}=300$. This scaling with $\gamma_{\rm sh}$ arises because the
distance between a suprathermal particle and the shock front increases
as $(1-\beta_{\rm sh})c\Delta t \simeq c\Delta t/(2\gamma_{\rm sh}^2)$
with time $\Delta t$ since shock crossing, and because the typical
residence time spent upstream (i.e. before return to the shock front)
$\Delta t\sim r_{\rm L,0}/\gamma_{\rm sh}$ (Achterberg et al. 2001,
Milosavljevic \& Nakar 2006, Pelletier et al. 2009).

\subsection{Microphysics of shock acceleration}

The above teaches us that the generation of turbulence in the shock
environment, the process of particle acceleration and the structure of
the shock itself form an inseparable tryptich, which must be
considered as a whole. As studies of non-relativistic magnetospheric
shocks have taught us, the reflection of a fraction of the incoming
(upstream) population on the shock front constitutes an essential
ingredient of the formation of the collisionless shock (e.g. Leroy
1983). In the shock precursor, where both the reflected particles and
the incoming plasma meet, dissipation and the build-up of an
electromagnetic barrier are initiated through micro-instabilities.

The investigation of ultra-relativistic collisionless shocks started
at the turn of 1990 with J. Arons and co-workers, in the case of high
magnetization, say $\sigma > 0.03$ (Hoshino \& Arons 1991, Hoshino et
al. 1992, Gallant et al. 1992). The results revealed the interesting
physics of such shocks: the loop of particles that are reflected by an
intense magnetic barrier triggers a synchrotron maser instability that
in turn radiates a coherent electromagnetic wave towards upstream as
well as electromagnetic waves towards downstream that are absorbed,
thereby heating the particle population. If the incoming plasma is
composed of electrons and protons, the upstream coherent waves exerts
a ponderomotive force on the electrons, which leads to wakefield
heating/acceleration. Whether heating or powerlaw acceleration occurs
is presently debated on the basis of recent particle-in-cell (PIC)
simulations, see Hoshino (2008) and Sironi \& Spitkovsky (2011a).

The physics of the shock wave differs at lower magnetization. The
breakthrough PIC simulations of an unmagnetized shock (no mean field,
i.e. $\sigma=0$) in a pair plasma with $\gamma_{\rm sh}\simeq20$,
conducted by Spitkovsky (2008) has validated the paradigm that
combines the three fundamental processes: the formation of a
collisionless relativistic shock front with reflected particles, the
generation of magnetic turbulence and the generation of a power law
distribution through Fermi process. Reflected particles run ahead of
the shock, initiate the growth of magnetic micro-turbulence through
streaming type instabilities, most notably the Weibel (filamentation)
branch; and, self-consistently, the spatial growth of the turbulence
produces a partial reflection of the incoming population, which
sustains the dynamics of the shock front, the generation of
turbulence, etc. The simulation also indicate the formation of a
powerlaw of accelerated particles, thanks to the scattering on the
turbulence that these particles self-excite.

Particle-in-cell simulations have by now become an invaluable tool for
addressing the complex non-linear microphysics of these relativistic
collisionless shock waves. Recent simulations have confirmed the onset
of Fermi acceleration in unmagnetized pair and electron-proton plasmas
(Martins et al. 2009) and indicated the absence of powerlaw
acceleration in (superluminal) highly magnetized shock waves with
$\sigma$ close to unity (Sironi \& Spitkovsky 2009, 2011a).

As powerful as they are, current PIC simulations cannot yet probe the
temporal or spatial scales that are associated to observed high energy
phenomena. For instance, the long simulations of Sironi \& Spitkvosky
(2009, 2011a) extend to $\sim 10^3\omega_{\rm pi}^{-1}$ (downstream
rest frame), i.e. a fraction of a second for a typical interstellar
environment, which remains small compared to the comoving dynamical
timescale of a gamma-ray burst external shock wave, $R/(\gamma_{\rm
  sh}c)\sim 10^4\,$sec. Interestingly, the longer simulations of
Keshet et al. (2009), $\sim10^4\omega_{\rm pi}^{-1}$, do not reveal a
stationary shock structure but point towards a magnetized turbulence
that increases in scale and strength with time. It is tempting to
relate, as these authors do, such behavior with the acceleration of
particles to progressively higher energies.

The above remarks and findings lead to the following picture for
relativistic Fermi acceleration. The Fermi process takes place when
micro-turbulence can be generated with $\delta B\gg B$ and $\lambda_B
\ll r_{\rm L}$, as discussed previously. Otherwise, the superluminal
configuration of the magnetized shock wave, even weakly magnetized,
halts prematurely the Fermi cycles. Assuming that the micro-turbulence
has been generated and survives behind the shock front, one can check
that the first generations of Fermi accelerated particles scatter on
this turbulence. To see this, write $\epsilon_B$ the fraction of shock
dissipated energy density $2\gamma_{\rm sh}^2n_{\rm u}mc^2$ (with
$m=m_p$ for an $e-p$ shock, $m=m_e$ for a pair shock) that has been
stored in magnetized microturbulence on typical scale
$\lambda_B$. Particle-in-cell simulations indicate $\epsilon_B\sim
10^{-2}$ and $\lambda_B\sim c/\omega_{\rm pi}$ for weakly magnetized
shocks. Then the ratio $r_{\rm L}/\lambda_B \sim \epsilon_B^{-1/2}
\gamma/\gamma_{\rm min}$, with $\gamma$ the Lorentz factor of the
particle and $\gamma_{\rm min}$ the Lorentz factor of shock heated
particles; for electrons, $\gamma_{\rm min}\sim \gamma_{\rm sh}
m_p/m_e$ if equipartition with the protons is reached, as PIC
simulations indicate (Sironi \& Spitkovsky 2011a). One then obtains
$r_{\rm L}\gtrsim \lambda_B$, meaning that the suprathermal particles
effectively probe the immediate vicinity of the shock. How the
turbulence evolves away from the shock remains an important open
question (Chang et al. 2008, Keshet et al. 2009), which may influence
the physics of acceleration to higher energies (Katz et al. 2007).

But, to accelerate particles to high energies, one first need to
ignite the Fermi process and this can be understood as follows. A
suprathermal particle carries a typical energy $\gamma_{\rm sh}^2
mc^2$, as measured in the upstream rest frame; let $\xi_{\rm cr}$
denote the fraction of shock dissipated energy carried by the
suprathermal population. For purely kinematic reasons, these particles
make up a sharp beam in phase space with small angular dispersion
$\lesssim 1/\gamma_{\rm sh}$ along the shock normal; indeed, particles
with angle $\gtrsim 1/\gamma_{\rm sh}$ travel along the shock normal
at a velocity $\lesssim \beta_{\rm sh}$ hence they cannot outrun the
shock (Gallant \& Achterberg 1999). This beam of particles carry a
plasma frequency $\omega_{\rm pb} \simeq \xi_{\rm
  cr}^{1/2}(m_e/m_p)^{1/2}\omega_{\rm pe}$, much smaller than the
background plasma frequency, even though the beam density is much
larger, $n_{\rm b}\simeq \gamma_{\rm sh}^2n_{\rm u}$. Through its
mixing with the background plasma particles, this beam induces
streaming instabilities. Various instabilities have been considered in
the literature, see Bret (2009), Lemoine \& Pelletier (2010), and Bret
et al. (2010) for a compilation of fast growing modes, but the fastest
appear to be the well studied filamentation instability (Medvedev \&
Loeb 1999, Wiersma \& Achterberg 2004, Lyubarsky \& Eichler 2006,
Achterberg \& Wiersma 2007, Achterberg et al. 2007, Lemoine \&
Pelletier 2010, 2011a, Rabinak et al. 2010) and the two stream
instability in its oblique version (Bret et al. 2005, Lemoine \&
Pelletier 2010, 2011a). In a weakly magnetized $e-p$ shock, the
electrons can be efficiently preheated to near equipartition in the
shock precursor (Lemoine \& Pelletier 2011a), in which case the
filamentation instability becomes the fast growing mode.

The unstable modes grow only if their growth timescale $\tau_{\rm
  inst.}$ is shorter than the time it takes for the precursor to
overtake the plasma element. In detail, this means $\tau_{\rm inst.} <
r_{\rm L,0}/(\gamma_{\rm sh}^3 c)$, given the precursor length scale
discussed earlier. For the filamentation instability, this condition
can be rewritten (see Lemoine \& Pelletier 2010):
\begin{equation}
\xi_{\rm cr}^{-1} \gamma_{\rm sh}^2 \sigma < 1\ .
\end{equation}
The higher the magnetization, or the larger the shock Lorentz factor,
the shorter the precursor, hence the lesser the amount of time for
microturbulence to grow.

\begin{figure}[ht]
\begin{center}
 \includegraphics[width=0.9\textwidth]{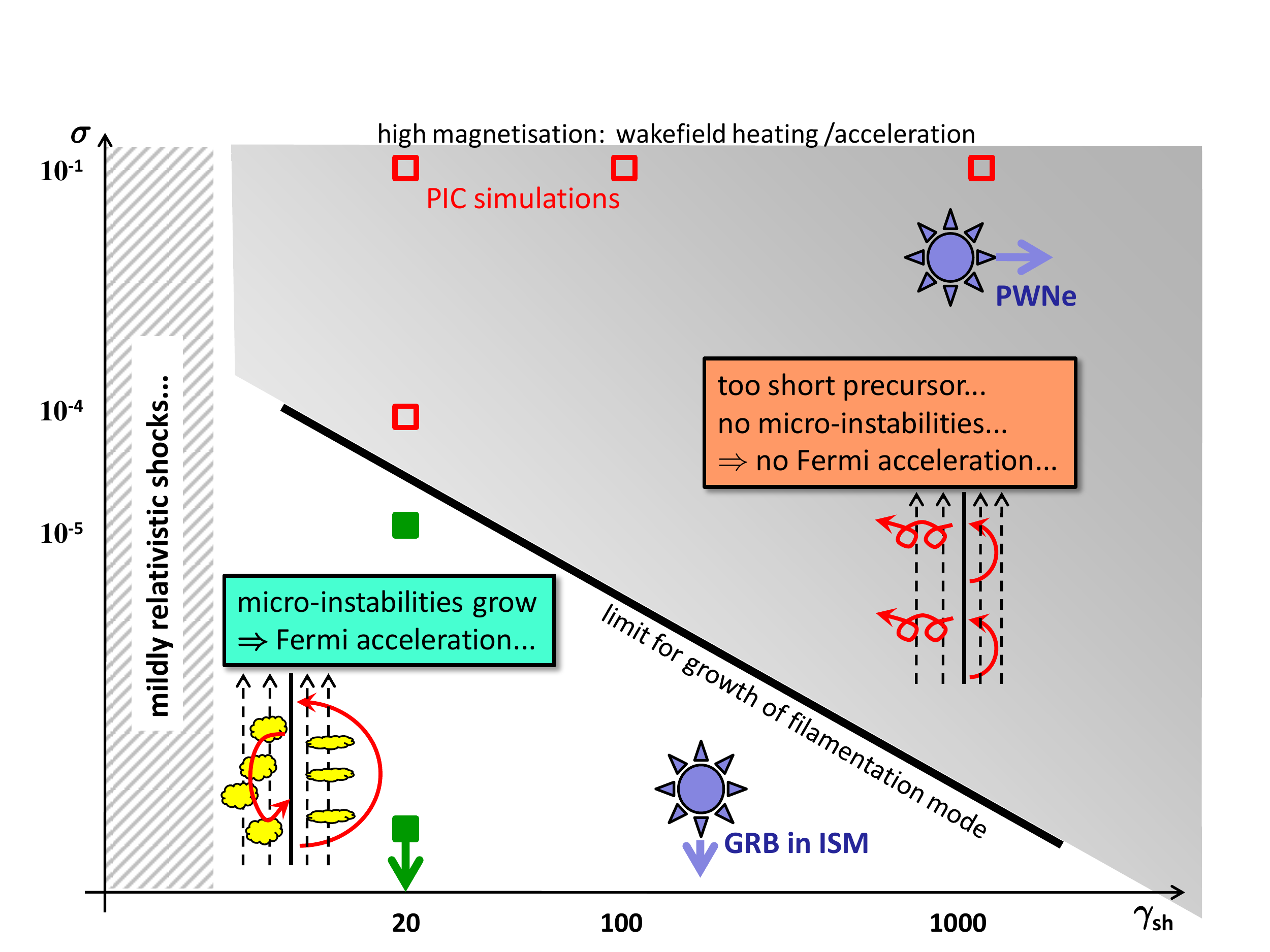}
  \caption{{\small Parameter space for relativistic shocks with shock Lorentz
    factor $\gamma_{\rm sh}$ in abscissae and magnetization of the
    incoming plasma $\sigma$ in ordinates. In the gray region, the
    precursor is too short to allow the growth of micro-instabilities
    by suprathermal particles, hence Fermi acceleration cannot take
    place (under the assumptions discussed in the text). The squares
    indicate the results of recent PIC simulations (Sironi \&
    Spitkovsky 2011a), which validate where applicable, this model:
    empty squares indicate no evidence for particle acceleration while
    filled squares mean that powerlaw Fermi type acceleration has been
    observed. The region at low $\gamma_{\rm sh}$ corresponding to
    mildly relativistic shocks is yet unexplored. See Lemoine \&
    Pelletier (2010) for a more detailed version of this figure.}}
\label{fig1}
\end{center}
\end{figure}

This leads to Fig.~\ref{fig1}, which delimits the region of parameter
space $\gamma_{\rm sh}$ (abscissae) -- $\sigma$ (ordinates) in which
the above condition is satisfied. This figure must be understood as a
qualitative representation and more quantitative details can be found
in Lemoine \& Pelletier (2010). The results of recent PIC simulations
(Sironi \& Spitkovsky 2011a) for $e-p$ shocks with $\gamma_{\rm
  sh}=20$ are represented as open (red) squares when no evidence for
particle acceleration has been observed in these simulations, or as
filled (green) squares if particle acceleration has been observed.
For this restricted set of simulations, the agreeement is therefore
highly satisfactory. In Fig.~\ref{fig1}, the limit has been taken at
$\xi_{\rm cr}^{-1}\gamma_{\rm sh}^2\sigma = 0.3$, corresponding to a
few efolds of growth of the turbulence on a precursor crossing
timescale. This figure also gives typical positions in this diagram
for gamma-ray burst external shock waves propagating in the
interstellar medium (annotated ``GRB in ISM'') and pulsar wind nebulae
(``PWNe''). These specific objects will be discussed in more detail in
the following.

Figure~\ref{fig1} omits the growth of Whistler waves in the
precursor. As discussed in Lemoine \& Pelletier (2010), such waves can
grow in the cold plasma limit for magnetization levels as high as
$\xi_{\rm cr}\gamma_{\rm sh}^{-3}m_p/m_e$, hence $\sigma\sim10^{-2}$
for $\gamma_{\rm sh}\sim20$, $\xi_{\rm cr}\sim0.1$. However, the
Whistler inertial range disappears progressively as electrons are
heated to relativistic temperatures, and the growth rate also
decreases. Sources of electron heating are discussed in Lemoine \&
Pelletier (2011a); they include, in particular, the oblique two stream
instability and the Buneman instability at the tip of the precursor.

To summarize, relativistic Fermi acceleration requires a weakly
magnetized upstream medium, all the more so at large Lorentz
factors. Accelerated particles scatter and possibly cool in the
turbulence that they themselves excite. Yet, many questions remain
open. To quote but a few: the fate of the turbulence in the post-shock
plasma, which controls the fate of acceleration to higher energies (as
higher energy particles probe longer distances in the post-shock
plasma), the dynamical evolution of the shock structure with time, and
the physics of acceleration in the mildly relativistic ($\gamma_{\rm
  sh}\sim 1-5$) regime, shown in cross-hatched in Fig.~\ref{fig1}.

The above discussion rests on a few specific assumptions. In
particular, one has neglected the radiative interactions of the
accelerated particles in the course of the Fermi cycles, one has
assumed the shock front to be steady, planar and one has also
neglected dissipation of the magnetic field in the shock transition,
or even an external source of turbulence. If either of these
assumptions were violated, one could expect interesting developments
with respect to particle acceleration, some of which are currently the
focus of interest. Consider for instance the possibility of radiative
interactions. As shown by Derishev et al. (2003), a charged particle
can be converted through such an interaction into a neutral state
which may then travel freely across the magnetic field. This neutral
particle may later be converted back into the charged state well ahead
of the shock front, thus opening Fermi cycles with widely different
particle kinematics. For example, an electron could upscatter a
photon, that would travel ahead of the shock, turn back through pair
production into an electron/positron, that would turn around in the
turbulence before encountering the shock front, while protons could be
turned into neutrons through photopion interactions, the neutron
decaying back into a proton ahead of the shock front... In this
``converter'' mechanism, the typical energy gain per cycle is much
larger than unity, possibly as large as $\gamma_{\rm sh}^2$, due to
the approximate isotropy of the suprathermal population. This may lead
to distorted powerlaw spectra with interesting phenomenology.

If the magnetic field is dissipated in the shock transition through
e.g., reconnection (Lyubarsky 2003), or sourced externally in the
shocked region, one may expect Fermi acceleration to become
operative. Although, in the latter case, stringent conditions apply; in
particular, the instability needs to grow faster behind the shock than
a typical Larmor time of the suprathermal particles in the background
magnetic field, otherwise these particles would not execute Fermi
cycles.

\section{Consequences and applications}

This Section discusses some applications and consequences of the above
microphysical view of relativistic shock acceleration for gamma-ray
bursts, pulsar wind nebulae and powerful active galactic nuclei.

\subsection{Acceleration in astrophysical sources}

\subsubsection{Gamma-ray bursts}

Regarding gamma-ray bursts, there exist strong evidence -- for a
substantial fraction of observed long gamma-ray bursts -- that the
afterglow emission is associated with synchrotron emission of shock
accelerated electrons with $\gamma_{\rm sh}\sim100$ at afterglow onset
(e.g. Piran 2005 for a review). As the shock wave sweeps up matter, it
decelerates and the external shock slowly transits towards the mildly
relativistic then non-relativistic regimes.

If the magnetization is low and the external medium composed of
electrons and protons, as one expects for an interstellar like
circumburst medium ($\sigma\sim10^{-9}$ for $B_{\rm u}\sim3\,\mu$G,
$n_{\rm u}\sim1\,$cm$^{-3}$), micro-instabilities develop efficiently
in the shock precursor and lead to efficient preheating of the
electrons and efficient Fermi acceleration.

Since the scattering occurs in short scale turbulence, one may expect
a radiative signature different from the standard synchrotron
paradigm. The physics is here characterized by the so-called wiggler
parameter $a \equiv e\delta B \lambda_B/ (m_ec^2)$, which measures the
capability of the magnetic force to deviate a relativistic electron of
Lorentz factor $\gamma$ by an angle $1/\gamma$ during the crossing of
a coherence length (e.g. Medvedev 2000, Fleishman \& Toptygin 2007,
Kirk \& Reville 2010). When $a >1$ the magnetic field produces a
single deviation of the electron in the emission cone of half angle
$1/ \gamma$, whereas when $a<1$ the electron can undergo several
wiggles in the emission cone.  For a relativistic shock, $a\sim
\xi_B^{1/2} \gamma_{\rm sh}m_p/m_e\gg1$, hence the emission
behaves as normal synchrotron radiation in a mean field at the peak
frequency, except that there is no polarization. A modification of the
spectral shape may however appear in the small frequency domain, see
e.g. (Medvedev 2006, Fleishman \& Urtiev 2010, Reville \& Kirk 2010).

Quite remarkably, there exists an almost universal energy limit when
the electron scatters and cools in intense small scale turbulence: by
comparing the scattering timescale $t_{\rm scatt}\sim
\gamma^2m_e^2c/(e^2\delta B^2 \lambda_B)$, which characterizes the
acceleration timescale, with the synchrotron cooling timescale $t_{\rm
  syn}=6\pi m_e c/(\delta B^2\sigma_{\rm T}\gamma)$, one can bound the
maximal Lorentz factor by above as (see also Kirk \& Reville 2010):
\begin{equation} 
  \gamma_{\rm max} \simeq \left(\frac{6\pi e^2 \lambda_B}{\sigma_{\rm T} m_e
      c^2}\right)^{1/3} 
  \approx \left(\frac{m_p/m_e}{n_{\rm u}r_e^3}\right)^{1/6} \sim
  10^6n_{\rm u,0}^{-1/6} \ ,
\end{equation}
where the second equality on the r.h.s. assumes
$\lambda_B=c/\omega_{\rm pi}$ and $r_e$ denotes the classical electron
radius.  The corresponding peak energy for the emitted photons reads
\begin{equation}
  \epsilon_{\gamma, \rm max} = \frac{9}{4\pi} \epsilon_B \gamma_{\rm sh}^2
  \frac{m_p c^2}{\gamma_{\rm max}\alpha_{\rm e.m.}} \approx 
  1\,{\rm GeV}\,\,
  \epsilon_{B,-2}^{1/2}\gamma_{\rm  sh,2.5}^2n_{\rm u,0}^{1/6}\ ,
\end{equation}
with the usual generic notation: $Q_x\equiv 10^{-x}Q $ in cgs.  Thus a
single synchrotron-like spectrum extending up to several GeV, even
possibly a few tens, can be expected. The performance of relativistic
shocks for electron acceleration and radiation is thus noteworthy. As
electrons cool faster than a hydrodynamical time in the early
afterglow, the conversion factor to radiation (synchrotron + inverse
Compton) is large $\sim \epsilon_e\sim 10\,$\%.  Let us note finally
that the inferred spectral index is generally in satisfactory agreement
with an expected $s\sim 2.2$ for relativistic shock acceleration with
isotropic scattering.

One must also consider the possibility that the outflow impinges on a
magnetized circumburst medium, either because the interstellar
magnetic field is relatively strong (meaning $B_{\rm u}\gtrsim 1\,$mG)
or because the explosion takes place in a magnetized stellar
wind. Recall that Li \& Waxman (2006) find $B\gtrsim 300\,\mu$G for
the {\em upstream} magnetic field during the early X-ray afterglow,
which either imply amplification/generation of the magnetic field by
streaming instabilities, as discussed earlier, or that the
pre-existing magnetic field already verifies that limit.  At such high
magnetization, Fermi acceleration may well be initially inhibited but
set on at later stages when $\gamma_{\rm sh}$ has dropped to
sufficiently low values. Such a scenario has been investigated in
Lemoine \& Pelletier (2011b), where it is found that: initially, one
records in the X-ray range the synchrotron emission of the thermal
shock heated electron population; but, as the peak frequency exits the
X-ray band on sub-day scales, one observes a drop-out in X-ray due to
the absence of Fermi powerlaws on timescales $\sim 100\,$s; finally,
at later times, e.g. $\gtrsim10^4\,$s, Fermi acceleration becomes
operative, powerlaws develop and electrons radiate in X-rays, so that
one recovers the standard afterglow light curve. This signature is
interesting because of its definite character but also because it may
explain some of the peculiar X-ray light curves recently observed by
the Swift satellite.

The above considerations apply to the ultra-relativistic external
shock. As mentioned previously, particle acceleration in the mildly
relativistic regime remains open for study. On naive grounds, one may
however expect these shocks to be very efficient particle accelerators
for the following reasons: superluminality is no longer generic, just
as shock compression of the magnetic field is less important; at a
given magnetization, the precursor extends to larger distances, which
opens the way for other instabilities, for instance resonant Whistler
wave excitation (Lemoine \& Pelletier 2010), or MHD modes; electron
heating to near equipartition remains envisageable, which would
circumvent the problem of electron injection in non-relativistic
shocks.

\subsubsection{Pulsar wind nebulae and active galactic nuclei}
The remarkable non-thermal spectrum of the Crab nebula indicates that
electrons are accelerated to energies as high as a PeV at the
termination shock of the relativistic pulsar wind, giving rise to
synchrotron radiation up to GeV energies, see e.g Kirk et al. (2009)
for a review. The changes in spectral slope indicate however that
possibly two different mechanisms are at work: a hard powerlaw with
$s\sim 1.5$ is observed up to TeV energies, and a  Fermi-like
powerlaw with $s\simeq 2.2$ is found above a TeV, up to the PeV.

The values of the Lorentz factor and the magnetization of the pulsar
wind at the termination shock are not resolved. Actually, these
quantities likely depend on the location on the termination shock
surface, which is far from planar (Komissarov \& Lyubarsky
2003). However its is argued that $\gamma_{\rm sh}\gtrsim 100$ at
least and $\sigma$ is not far below unity (Kirk et al. 2009). The
model developed in the previous Section then indicates that Fermi
acceleration should not be operative in such conditions. One
possibility of course, is that wakefield acceleration of electrons and
positrons occurs in the shock precursor if sufficiently many ions are
present, as proposed by Hoshino et al. (1992), see however Kirk et
al. (2009).

Given the rather nice agreement of the spectral slope expected for
Fermi acceleration with that observed above a TeV, one may turn the
question otherwise and ask what would be needed to allow efficient
Fermi acceleration to the highest energies. One interesting
possibility is that shock driven reconnection of the striped wind
leads to magnetic dissipation and electron/positron acceleration with
the hard powerlaw index up to TeV (Lyubarsky 2003, P\'etri \&
Lyubarsky 2007, Sironi \& Spitkovsky 2011b). Results from
particle-in-cell simulations are encouraging in this respect. How
Fermi acceleration emerges at higher energies is not yet well
understood, although one may speculate that the dissipation of
magnetic field in the shock transition suffices to remove the phase
space locking on magnetic field lines that was discussed earlier or,
that reconnection accelerated electrons at a TeV -- that carry most of
the energy -- propagate sufficiently far upstream to induce
instabilities, that then provide a seed for scattering of higher
energy particles.

Finally, in the case of blazars, Bl Lac objects or flat spectrum radio
quasars, i.e. relativistic jets of active galactic nuclei that are
beamed towards the observer, it is not known where the acceleration
process takes place, whether in mildly relativistic internal shocks
that are embedded in a relativistic flow, as for the prompt emission
of gamma-ray bursts, or at the external shock for instance. In any
case, recent inferences of hard powerlaws from blazar spectra at very
high energies point to interesting acceleration physics in
these objects (e.g. Ackermann et al. 2011).

\subsection{Acceleration to ultra-high energies}

A well known estimate of the maximal acceleration energy for
non-relativistic shocks compares the acceleration timescale, $t_{\rm
  acc} \sim t_{\rm scatt}/\beta_{\rm sh}^2$ (with $t_{\rm scatt}\sim
D/c^2$ the diffusion timescale in the shock environment) to the age of
the shock wave $R/(\beta_{\rm sh}c)$ to infer (Lagage \& C\'esarsky
1983, Hillas 1984): $E_{\rm max} \sim \beta_{\rm sh} e R B$ if Bohm
diffusion applies in the magnetic field $B$, meaning $t_{\rm
  scatt}\sim t_{\rm L}$. This estimate suggests that relativistic
shock waves with $\beta_{\rm sh}\rightarrow 1$ appear as efficient
particle accelerators to ultra-high energy.

However, if the scattering takes place in micro-turbulence, the
scattering timescale increases quadratically with the energy $E$,
since $t_{\rm scatt} \sim t_{\rm L}^2c/\lambda_B$. In the presence of
a mean magnetic field, one may argue that the upstream residence time
is either governed by the scattering in the micro-turbulence,
i.e. $t_{\rm u} \sim E^2/(\gamma_{\rm sh}^2e^2 \delta B^2 c
\lambda_{B,\rm u})$ or by the rotation of the particles in the
background magnetic field $t_{\rm u}\sim E/(\gamma_{\rm sh}e B_{\rm u}
c)$ (e.g. Pelletier et al. 2009). A Bohm estimate in the
self-generated magnetic field is thus not supported by theory in the
ultra-relativistic limit. In either limit above, ultra-relativistic
shocks do not appear as promising accelerators to reach
ultra-energies. For a quantitative estimate, consider the external
shock of a gamma-ray burst at radius $R\sim 10^{17}\,$cm, assume that
transport is governed by scattering in the upstream micro-turbulent
field $\delta B_{\vert\rm u}$, assume that the acceleration timescale
reduces to the residence time (a conservative bound), then the maximal
electron energy, notwithstanding possible energy losses:
\begin{equation}
  E_{\rm max} \simeq \gamma_{\rm sh}\sqrt{R\lambda_B}Z e \delta
  B_{\vert\rm u}c \sim 10^7\,{\rm GeV}\,\, Z
  \epsilon_B^{1/2}R_{17}n_{\rm u,0}\ .
\end{equation}

Mildly relativistic shocks appear as more promising sources of
ultra-high energy particles. A reasonable guess is to expect a
substantial magnetic amplification at such shocks with efficiency
$\xi_B\sim 1-10$\%, on rather large MHD scales. Such a configuration
might possibly lead to a Bohm regime for particle transport. Then one
would reach an optimal acceleration timescale with $t_{\rm acc} \sim
t_{\rm L}$ and maximal energy $E_{\rm max}\sim Z e \delta B$ (as
written in the upstream rest frame).

To provide another perspective on this issue, one may express the
acceleration timescale in units of the Larmor time in the total
magnetic field, i.e. $t_{\rm acc}={\cal A}t_{\rm L}$ and relate the
maximal energy to the magnetic luminosity of the source (e.g. Norman
et al. 1995, Waxman 2005, Lemoine \& Waxman 2009). Assuming that the
acceleration site resides in a relativistic outflow with bulk Lorentz
factor $\gamma$ (velocity $\beta c$), beamed towards the observer,
with half-opening angle $\Theta$, limiting the maximal energy by the
dynamical timescale as previously, one finds that the source frame
magnetic luminosity of the source $L_B\,=\,R^2\Theta^2 \gamma^2\beta c
B^2/4$ must verify
\begin{equation}
L_B\,\geq\, 0.65\times 10^{45}\, \Theta^2\gamma^2{\cal
  A}^2\beta^3Z^{-2}E_{20}^2\, {\rm erg/s}\ ,\label{eq:LB}
\end{equation}
in order to produce $10^{20}E_{20}\,$eV particles (of charge $Z$) in
the observer rest frame. This bound generalizes the previous Lagage \&
C\'esarsky (1983) limit to relativistic outflows. It is quite
remarkable to note that only a handful of relativistic sources appear
capable of accelerating particles to the most extreme energies
recorded if $Z\sim 1$. In particular, as discussed in Lemoine \&
Waxman (2009), this excludes standard active galactic nuclei or even
blazars if the accelerated species are protons, while jets of FR-I
radio-galaxies might possibly accelerate heavier particles to $Z\times
10^{18}\,$eV. Whether the most extreme cosmic rays are protons or
irons is a current subject of controversy, which clearly influences
the debate on the nature of their sources. Such considerations lie
however beyond the scope of this review.

\section{Conclusion}
The three facets of a collisionless shock structure, which include a
reflecting barrier for a part of the incoming particles, the
generation of supra-thermal particles and the generation of magnetic
turbulence altogether frame a successful paradigm that applies to
astrophysical shocks, both non-relativistic and
relativistic. Although, the phenomenology in the ultra-relativistic
shock limit differs appreciably from that in the non-relativistic
limit as the shock front then moves at a velocity close to $c$, about
as fast as the accelerated particle. While in non-relativistic shocks,
suprathermal particles isotropize and diffuse in the magnetized
turbulence on both sides of the shock front, the suprathermal
population around relativistic shocks remains highly anisotropic. In
non-relativistic shocks, the small energy gain per cycle around the
shock wave is compensated by a small escape probability through
advection with the shocked plasma, while in relativistic shocks, this
probability is high but the energy gain is also substantial. While the
precursor of non-relativistic shocks extends to very large scales,
that of relativistic shocks lies on scales intermediate between the
plasma skin depth scale and the Larmor scale of the cycling
particles. Finally, while perpendicular shock acceleration may be
efficient at non-relativistic velocities, ultra-relativistic shocks
are generically perpendicular and for them, acceleration is prohibited
unless small scale turbulence has been excited in the shock vicinity.

The above picture has by now been understood analytically and tested,
in part, by sophisticated particle-in-cell simulations. One thus
understands why Fermi acceleration should not occur at
ultra-relativistic shocks with magnetization of order unity, a rather
frequent situation in high energy astrophysics: the larger the Lorentz
factor, or the larger the magnetization, the shorter the precursor,
hence the lesser the amount of time available for excitation of short
scale turbulence by reflected particles. As we have argued here, the
dividing line appears to be $\xi_{\rm cr}^{-1}\sigma\gamma_{\rm
  sh}^2\sim1$: for larger values of the l.h.s., growth of the
filamentation instability and of other microturbulent modes, is
suppressed.

Of course, additional effects may help acceleration proceed in a
magnetized environment. Among these, one may cite the wakefield
acceleration of electrons (for an $e-p$ shock) in precursor waves
emitted by the incoming protons, the acceleration of particles in
shock driven reconnection of an alternating upstream magnetic field,
or radiation driven acceleration through repeated charged to neutral
state conversions. And, of course, many questions remain open, among
which: how are electrons heated in the shock transition and
whether/how they reach equipartition with the protons, how do the
electrons radiate and whether the radiative signature may provide a
diagnosis of the turbulent state, whether the shock front remains
stationary or whether it may suffer from corrugation or reformation,
how acceleration proceeds at mildly relativistic shocks,  etc.
\bigskip

{\noindent\bf Acknowledgments}:
We acknowledge support from the GDR-PCHE and the PEPS/PTI program of CNRS/INP.
\bigskip


\begin{thebibliography}{99}

\bibitem{} A. Achterberg, Y.  Gallant, J. G. Kirk, A. W. Guthmann,
  A. W., MNRAS 328, 393-408 (2001)

\bibitem{} A. Achterberg, J.  Wiersma, Astron. Astrophys., 475, 19-36 (2007)

\bibitem{} A. Achterberg, J. Wiersma, C. A. Norman,
  Astron. Astrophys., 475, 1-18 (2007)

\bibitem{} M. Ackermann et al. (Fermi Collaboration),
  arXiv:1108.1420 (2011)

\bibitem{} J. Bednarz, M. Ostrowski, Phys. Rev. Lett., 80, 3911-3914 (1998)

\bibitem{} M. C. Begelman, J. G. Kirk, ApJ, 353, 66-80 (1990)

\bibitem{} A. Bell, MNRAS, 182, 147-156 (1978)

\bibitem{} R. Blandford, C. McKee, Phys. Fluids, 19, 1130-1138 (1976)

\bibitem{} A. Bret, M.-C. Firpo, C. Deutsch, Phys. Rev. Lett., 94,
  115002 (2005a)

\bibitem{} A. Bret, M.-C. Firpo, C. Deutsch, Phys. Rev. E,
  72, 016403 (2005b)
 
\bibitem{} A. Bret, ApJ, 699, 990-1003  (2009)

\bibitem{} A. Bret, L. Gr\'emillet, D. B\'enisti, Phys. Rev. E, 81,
  036402 (2010)

\bibitem{} P. Chang, A. Spitkovsky, J.  Arons, ApJ, 674, 378-387 (2008)

\bibitem{} E. V. Derishev, F. A. Aharonian, V. V. Kocharovsky, Vl. V.,
  Kocharovsky, Phys. Rev. D, 68, 043003 (2003)

\bibitem{} D. Ellison, G. Double, Astropart. Phys., 18, 213-228 (2002)

\bibitem{} G. D. Fleishman, I. N. Toptygin, MNRAS, 381, 1473-1481
  (2007).

\bibitem{} G. D. Fleishman, F. A. Urtiev, MNRAS, 406, 644-655
  (2010).

\bibitem{} Y. A. Gallant, A. Achterberg, MNRAS, 305,
  L6-L10 (1999)

\bibitem{} Y. A. Gallant, M. Hoshino, A. B. Langdon, J. Arons,
  C. E. Max, ApJ, 391, 73-101 (1992)

\bibitem{} A. Gruzinov, E. Waxman, ApJ, 511, 852-861 (1999)

\bibitem{} A. M. Hillas,  Ann. Rev. Astron. Astrophys., 22, 425-444
  (1984).

\bibitem{} M. Hoshino, ApJ, 672, 940-956 (2008)

\bibitem{} M. Hoshino, J. Arons, Phys. Fluids B, 3, 818-833 (1991)

\bibitem{} M. Hoshino, J. Arons, Y. A. Gallant, A. B. Langdon, ApJ,
  390, 454-479 (1992).

\bibitem{} B. Katz, U. Keshet, E. Waxman, ApJ, 655, 375-390 (2007)
 
\bibitem{} U. Keshet, Phys. Rev. Lett., 97, 221104 (2006)
 
\bibitem{} U. Keshet, B. Katz, A. Spitkovsky, E. Waxman, E., 
  ApJ, 693, L127 (2009)

\bibitem{} U. Keshet, E. Waxman, Phys. Rev. Lett., 94,
  111102 (2005)

\bibitem{} J. G. Kirk, A. W. Guthmann, Y. A. Gallant, A. Achterberg,
  ApJ, 542, 235-242 (2000)

\bibitem{} J. G. Kirk, Y. Lyubarsky, J. P\'etri, in ``Neutron stars
  and pulsars'', Astrophysics and Space Science Library, 357,
  Springer, Berlin Heidelberg, 2009, p.421.

\bibitem{} J. G. Kirk, B. Reville, ApJ, 710, L16-L20 (2010)

\bibitem{} S. S. Komissarov, Y. E. Lyubarsky, MNRAS, 344, L93-L96 (2003)

\bibitem{} P.-O. Lagage, C. C\'esarsky, Astron. Astrophys., 125,
  249-257 (1983).

\bibitem{} M. Lemoine, G.  Pelletier, ApJ, 589, L73-L76 (2003)

\bibitem{} M. Lemoine, G. Pelletier, MNRAS, 402, 321-334 (2010)

\bibitem{} M. Lemoine, G. Pelletier, MNRAS, 417, 1148-1161 (2011a)

\bibitem{} M. Lemoine, G. Pelletier, MNRAS, to appear, arXiv:1103.4823 (2011b)

\bibitem{} M. Lemoine, G. Pelletier, B. Revenu, ApJ, 645, L129-L132 (2006)

\bibitem{} M. Lemoine, B.  Revenu, MNRAS, 366, 635-644 (2006)

\bibitem{} M. Lemoine, E. Waxman, 2009, JCAP, 11, 009

\bibitem{} M. M. Leroy, Phys. Fluids, 26, 2742-2753 (1983).

\bibitem{} Z. Li, E. Waxman, ApJ, 651, 328-332 (2006)

\bibitem{} Z. Li, X.-H. Zhao, JCAP, 05, 008 (2011)

\bibitem{} Y. Lyubarsky, MNRAS, 345, 153-160 (2003).

\bibitem{} Y. Lyubarsky, D. Eichler, ApJ, 647, L1250-L1254 (2006)

\bibitem{} S. F. Martins, R. A. Fonseca, L. O. Silva, W. B. Mori,
  ApJ, 695, L189-L193 (2009).

\bibitem{} M. V. Medvedev, ApJ, 540, 704-714 (2000)

\bibitem{} M. V. Medvedev, ApJ, 637, 869-872 (2006).

\bibitem{} M. V. Medvedev, A. Loeb, ApJ, 526, 697-706 (1999)

\bibitem{} M. Milosavljevi\'c, E. Nakar, ApJ, 651, 979-984  (2006)

\bibitem{} J. Niemiec, M. Ostrowski, ApJ, 641, 984-992 (2006)

\bibitem{} J. Niemec, M. Ostrowski, M. Pohl, ApJ, 650, 1020-1027 (2006)

\bibitem{} C. A. Norman, D. B. Melrose, A. Achterberg, ApJ, 454, 60-68
  (1995).

\bibitem{} G. Pelletier, M. Lemoine, A. Marcowith, MNRAS,
  393, 587-597 (2009)

\bibitem{} J. P\' etri, Y. Lyubarsky, Astron. Astrophys., 473, 683-700 (2007)

\bibitem{} T. Piran, Rev. Mod. Phys., 76, 1143-1210 (2005)

\bibitem{} I. Rabinak, B. Katz, E. Waxman, ApJ, 736, 157-164 (2011)

\bibitem{} L. Sironi, A. Spitkovsky, ApJ, 698, 1523-1549 (2009)

\bibitem{} L. Sironi, A. Spitkovsky, ApJ, 726, 75-99 (2011a)

\bibitem{} L. Sironi, A. Spitkovsky, arXiv:1107.0977v1 (2011b)

\bibitem{} A. Spitkovsky, ApJ 682, L5-L8 (2008) 

\bibitem{} M. Vietri, ApJ, 591, 954-961 (2003).

\bibitem{} E. Waxman, Physica Scripta, T121, 147-152 (2005).

\bibitem{} J. Wiersma, A. Achterberg, Astron. Astrophys., 428,
  365-371 (2004)

\end{thebibliography}
\end{document}